\documentclass{PoS}
\usepackage[utf8]{inputenc}
\usepackage{url}
\usepackage{amsmath}
\usepackage{xcolor}
\usepackage{cite}

\newcommand\sqrts{$\sqrt{s_\mathrm{NN}}$ }
\newcommand\smsh{\texttt{SMASH} }
\newcommand\hybrid{\texttt{SMASH-vHLLE-hybrid} }
\newcommand\vhlle{\texttt{vHLLE} }

\newcommand\pt{$p_\mathrm{T}$ }

\newcommand\pim{$\pi^-$}
\newcommand\Km{$K^-$}

\title{Conservation laws in a novel hybrid approach}

\ShortTitle{Conservation laws in a novel hybrid approach}

\author{\speaker{Anna Sch\"afer} \\
                 Institut f\"{u}r Theoretische Physik,
                 Goethe Universit\"{a}t, Frankfurt am Main, Germany \\
                 Frankfurt Institute for Advanced Studies, Frankfurt am Main, Germany \\
                 GSI Helmholtzzentrum f\"{u}r Schwerionenforschung, Darmstadt, Germany \\
                 E-mail: \email{aschaefer@fias.uni-frankfurt.de}}
\author{Iurii Karpenko\\
        Faculty of Nuclear Sciences and Physical Engineering,
        Czech Technical University in Prague, B\v{r}ehov\'{a} 7, 11519 Prague 1, Czech Republic \\
        E-mail: \email{iurii.karpenko@fjfi.cvut.cz}}
\author{Hannah Elfner\\
        GSI Helmholtzzentrum f\"{u}r Schwerionenforschung, Darmstadt, Germany \\
        Institut f\"{u}r Theoretische Physik,
        Goethe Universit\"{a}t, Frankfurt am Main, Germany \\
        Frankfurt Institute for Advanced Studies, Frankfurt am Main, Germany \\
        Helmholtz Research Academy Hesse for FAIR (HFHF), GSI Helmholtz Center, Frankfurt am Main, Germany \\
        E-mail: \email{elfner@fias.uni-frankfurt.de}}

\abstract{Heavy-ion collisions covering a wide range of collision energies provide a vast amount of observables characterizing the properties of strongly-interacting matter. In particular collisions towards the high baryon-density regime of the QCD phase-diagram have become of interest to study the postulated first order phase transition and to locate a possible critical end point.
In this work, the \hybrid is presented as a novel hybrid model to theoretically describe such heavy-ion collisions. In addition, the \smsh hadron resonance gas equation of state is introduced. The accuracy of the latter is shown to be of fundamental importance in order to conserve energy, baryon number and electric charge throughout the different stages of the hybrid model. Furthermore, the impact of an inaccurate equation of state on final state observables is discussed. This work constitutes a first validation of the \hybrid in terms of conservation laws and excitation functions. It is expected to be applied to a broader range of observables in the future.
}

\FullConference{CPOD2021 - International Conference on Critical Point and Onset of Deconfinement\\
                March 2021\\
                Virtual}

\begin{document}

\section{Introduction}
Heavy-ion collisions provide a unique opportunity to experimentally access QCD matter under extreme conditions, that is high temperatures and/or baryon densities. Particularly the properties of strongly-interacting matter at finite baryon densities, which are accessible through heavy-ion collisions at low and intermediate energies, have become of interest in recent years to study a postulated first oder phase transition and critical end point in the QCD phase diagram \cite{Stephanov:1998dy}. Dedicated heavy-ion programs such as the BESII program at BNL, the NA61/SHINE experiment at CERN, FAIR at GSI or NICA at JINR have already or will soon provide further experimental data in the designated region of interest \cite{Friese:2019qyw}. Theoretically, heavy-ion collisions at high energies are successfully described by hydrodynamics+transport models, while for those at low energies a pure transport description constitutes the standard approach \cite{Petersen:2017jdb}. For intermediate collision energies however, there is no such standard description yet, but dynamically initialized hybrid approaches are a promising candidate \cite{Akamatsu:2018olk}. In this work we introduce a novel hybrid approach consisting of the state-of-the-art hadronic transport model \smsh \cite{Weil:2016zrk, SMASH_github} and the 3+1D viscous hydrodynamics code \vhlle \cite{Karpenko:2013wva}.
The \hybrid was already successfully applied to study \mbox{(anti-)proton} annihilation and regeneration in heavy-ion collisions \cite{Garcia-Montero:2021haa}. In this study we further introduce the equation of state of the \smsh hadron resonance gas down to an energy density of $e = 0.01$ GeV/fm$^3$ and demonstrate the importance of its accuracy in the context of conservation laws. \\
This work is structured as follows: In Sec.~\ref{sec:model} the \hybrid is briefly decribed, with an introduction to the \smsh hadron resonance gas equation of state in Sec.~\ref{sec:model_eos}. Sec.~\ref{sec:results} contains an extensive discussion of quantum number conservation in the \hybrid as well as its implications for final state observables. A brief summary and outlook are provided in Sec.~\ref{sec:conclusions}.

\section{Model Description}
\label{sec:model}
The \hybrid \cite{hybrid_github} is a novel modular hybrid approach consisting of the hadronic transport approach \smsh \cite{Weil:2016zrk, SMASH_github, SMASH_doi}, the 3+1D viscous hydrodnamics code \vhlle \cite{Karpenko:2013wva}, and the \texttt{SMASH-hadron-sampler} \cite{Karpenko:2015xea} \footnote{For this work, \texttt{SMASH-2.0.2}, \texttt{vHLLE:bce38e0}, and \texttt{SMASH-hadron-sampler-1.0} are used.}.
It can be applied to describe heavy-ion collisions ranging from \sqrts = 4.3 GeV to \sqrts = 5.02 TeV. \\
In the \texttt{SMASH-vHLLE-hybrid}, the initial state is modeled by means of \smsh and extracted on a hypersurface of constant proper time. This proper time is determined from nuclear overlap, i.e. the passing time of the two nuclei, but is lower bound by $\tau_0 = 0.5$. Gaussian smearing, c.f. Table~\ref{tab:parameters}, is applied to provide smooth initial conditions for the event-by-event hydrodynamic evolution \cite{Karpenko:2015xea}. The latter is performed relying on a chiral model equation of state \cite{Steinheimer:2010ib} and utilizing the viscosities and smearing parameters listed in Table \ref{tab:parameters}.
\begin{table}
  \centering
  \begin{tabular}{ c | c c c c c c c c c c}
    \sqrts [GeV] & 4.3 & 6.4 & 7.7 & 8.8 & 17.3 & 27.0 & 39.0 &  62.4 & 130.0 & 200.0 \\
    \hline
    $\eta/s$ & 0.2 & 0.2 & 0.2 & 0.2 & 0.15 & 0.12 & 0.08 & 0.08 & 0.08 & 0.08 \\
    $R_\perp$ & 1.4 & 1.4 & 1.4 & 1.4 & 1.4 & 1.0 & 1.0 & 1.0 & 1.0 & 1.0 \\
    $R_\eta$ & 1.3 & 1.2 & 1.2 & 1.0 & 0.7 & 0.4 & 0.3 & 0.6 & 0.8 & 1.0
  \end{tabular}
  \caption{Shear viscosities ($\eta / s$), transverse Gaussian smearing parameters ($R_\perp$), and longitudinal Gaussian smearing parameters ($R_\eta$) applied in this work for the hydrodynamical evolution of different collision energies. Note, that Pb + Pb collisions are simulated at \sqrts = 6.4, 8.8, 17.3 GeV, Au + Au collisions at all other energies.}
  \label{tab:parameters}
\end{table}
The medium is evolved until reaching a hypersurface of constant energy density $e_\mathrm{crit} = 0.5$ GeV/fm$^3$, determined with the \texttt{CORNELIUS} subroutine \cite{Huovinen:2012is}. Particlization of the fluid elements, according to the \smsh hadron resonance gas, is achieved with the \texttt{SMASH-hadron-sampler} that effectively evaluates the Cooper-Frye formula \cite{Cooper:1974mv}. The sampled particles are further evolved in \texttt{SMASH} and the final interactions performed.
Note that the thermodynamic properties of the fluid elements on the freezeout surface need to match the \smsh equation of state, to assure  conservation of quantum numbers across the sampling process. This equation of state needs thus be determined from the \smsh hadron resonance gas to serve as supplementary input for the hydrodynamic evolution.

\subsection{The \smsh Hadron Resonance Gas Equation of State}
\label{sec:model_eos}
\begin{figure}[t]
  \centering
  \includegraphics[width=0.45\textwidth]{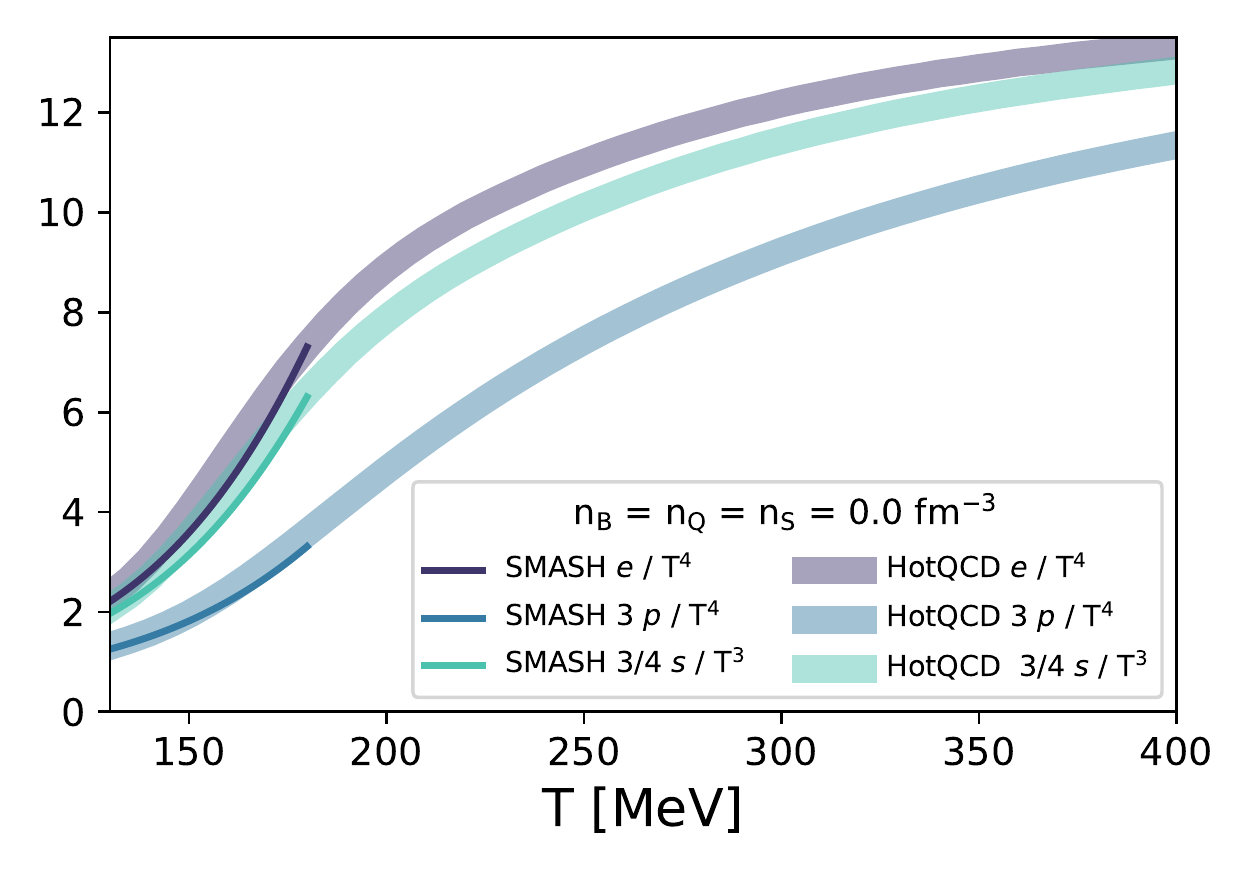}
  \caption{Energy density ($e$), pressure ($p$) and entropy density ($s$) from the \smsh hadron resonance gas equation of state (lines) in comparison to results from (2+1)-flavour lattice QCD by the HotQCD collaboration (bands) \cite{HotQCD:2014kol}.}
  \label{fig:EoS_comp}
\end{figure}

The \smsh hadron resonance gas consists of all hadrons listed by the PDG \cite{ParticleDataGroup:2020ssz} up to a mass of 2.35  GeV. These degrees of freedom occupy the gas of hadrons at given energy density $e$, baryon density $n_\mathrm{B}$, charge density $n_\mathrm{Q}$, and strangeness density $n_\mathrm{S}$ according to their quantum numbers.
As a consequence, they define the corresponding thermodynamic quantities temperature $T$, pressure $p$, baryon chemical potential $\mu_\mathrm{B}$, charge chemical potential $\mu_\mathrm{Q}$, and strangeness chemical potential $\mu_\mathrm{S}$. The equation of state generally contains the mapping $(e, n_\mathrm{B}, n_\mathrm{Q}, n_\mathrm{S}) \to (T, p, \mu_\mathrm{B}, \mu_\mathrm{S}, \mu_\mathrm{Q})$ encapsulating the properties of the underlying gas of particles.
For the determination of the \smsh equation of state state however, we neglect the explicit dependence on the strangeness density $n_\mathrm{S}$, for it can be approximated by $n_\mathrm{S} = 0$ fm$^{-3}$ in the context of heavy-ion collisions. The \smsh equation of state introduced in this work thus provides the mapping
\begin{equation}
  (e, n_\mathrm{B}, n_\mathrm{Q}) \ \to \ (T, p, \mu_\mathrm{B}, \mu_\mathrm{S}, \mu_\mathrm{Q}),
  \label{eq:EoS_map}
\end{equation}
It can be extracted by solving the set of coupled equations
\begin{equation}
  \label{eq:EoS_coupled}
  \begin{split}
    T &= T(e, n_\mathrm{B}, n_\mathrm{Q}) \\
    \mu_\mathrm{B} &= \mu_\mathrm{B}(e, n_\mathrm{B}, n_\mathrm{Q}) \\
    \mu_\mathrm{Q} &= \mu_\mathrm{Q}(e, n_\mathrm{B}, n_\mathrm{Q}) \\
    \mu_\mathrm{S} &= \mu_\mathrm{S}(e, n_\mathrm{B}, n_\mathrm{Q}),
  \end{split}
\end{equation}
where $e, n_\mathrm{B}$ and $n_\mathrm{Q}$ in turn depend on $T, \mu_\mathrm{B}, \mu_\mathrm{Q}$ and $\mu_\mathrm{S}$. The solutions of Eqs. (\ref{eq:EoS_coupled}) can in principle be determined numerically with a root solver. This solver is however highly-sensitive to the choice of the initial approximation and sometimes fails to converge. Especially for low energy densities ($e$ < 0.1 GeV/fm$^3$) and combinations of ($e, n_\mathrm{B}$) and ($e, n_\mathrm{Q}$) close to  kinematic thresholds\footnote{The kinematic thresholds for $n_\mathrm{B}$ and $n_\mathrm{Q}$ stem from the composition of the gas. In the case of the SMASH hadron resonance gas, the lightest baryon is the (anti-)proton with $m_\mathrm{p, \bar{p}} = 0.938$ GeV and the lightest charged particle is the pion with $m_\mathrm{\pi} = 0.138$ GeV. The physical region is thus restricted to $e \geq m_\mathrm{p} \ |n_\mathrm{B}|$ and $e \geq m_\pi \ |n_\mathrm{Q}|$.},
it is challenging to obtain a reliable result.
As a consequence, the herein presented equation of state of the \smsh hadron resonance gas is perfectly accurate for high $e$ as well as low $n_\mathrm{B}$ and $n_\mathrm{Q}$. In the problematic regions however, it constitutes an approximation\footnote{This approximation is obtained from varying grid spacings and initial guesses when solving Eqs.~(\ref{eq:EoS_coupled}) in combination with averages over different interpolations based on those solver results that are reliable.}. The equation of state is cut below $e = 0.01$ GeV/fm$^3$, for lack of reliable solutions. \\
To validate the resulting \smsh equation of state, the temperature, pressure and entropy density are compared to results from lattice QCD  \cite{HotQCD:2014kol} in Fig.~\ref{fig:EoS_comp}. Bands denote results from the HotQCD collaboration in (2+1)-flavour QCD and lines the \smsh hadron resonance gas.
A decent agreement, similar to that presented in Fig.~5 of \cite{HotQCD:2014kol}, is obtained in the low temperature regime.\\

In the following, two different kinds of the \smsh equation of state are used to demonstrate the effect of small inaccuracies in the equation of state on the conservation of quantum numbers as well as final state observables. The first equation of state relies solely on the root solver results for Eqs.~(\ref{eq:EoS_coupled}), without further modification, and is thus denoted the \emph{unmodified equation of state}. The second version of the equation of state contains the approximation in the problematic regions mentioned above and is denoted the \emph{improved equation of state}.

\section{Results}
\label{sec:results}
\begin{figure}[t]
  \centering
  \includegraphics[width=0.29\textwidth]{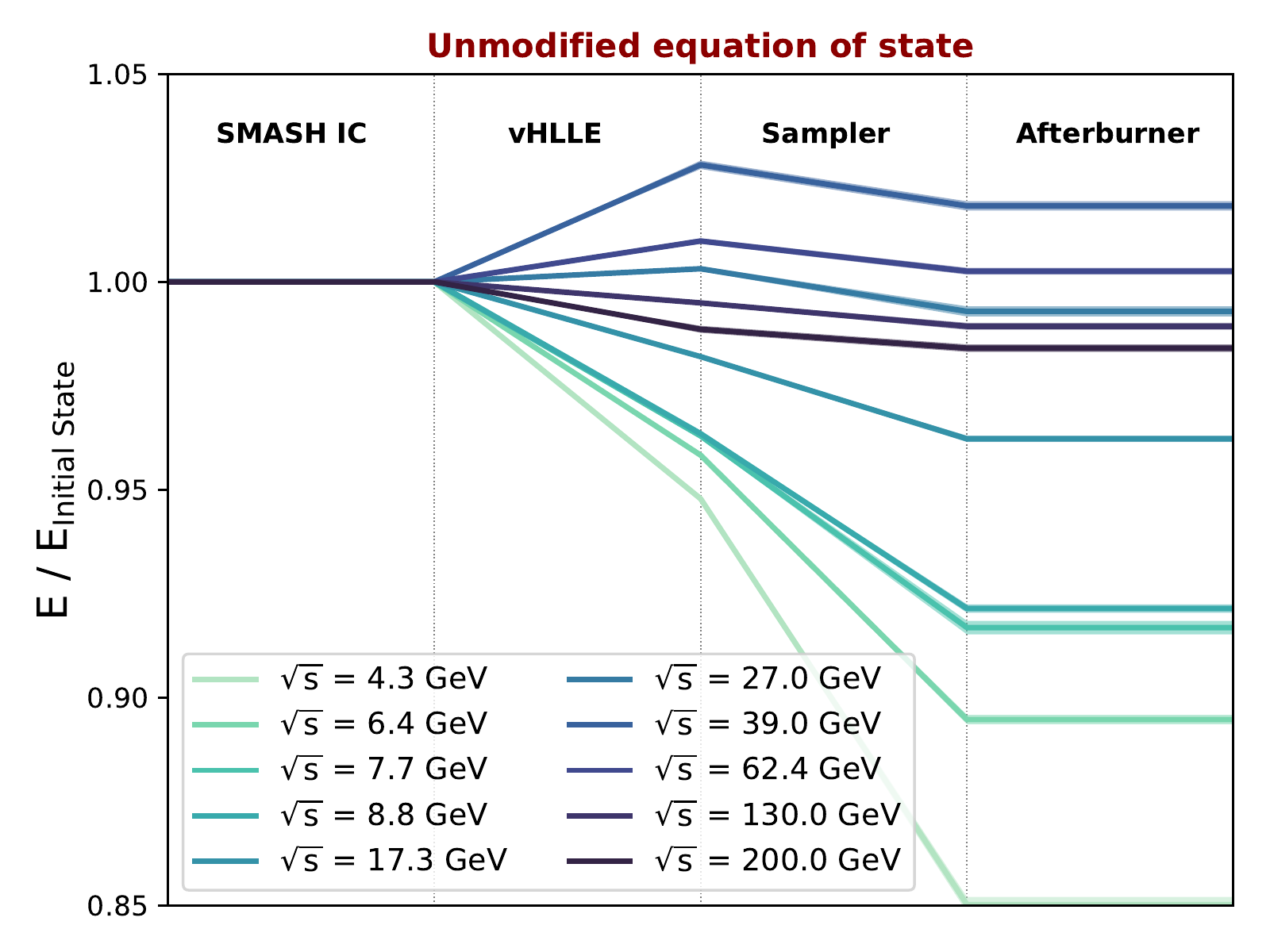}
  \hfill
  \includegraphics[width=0.29\textwidth]{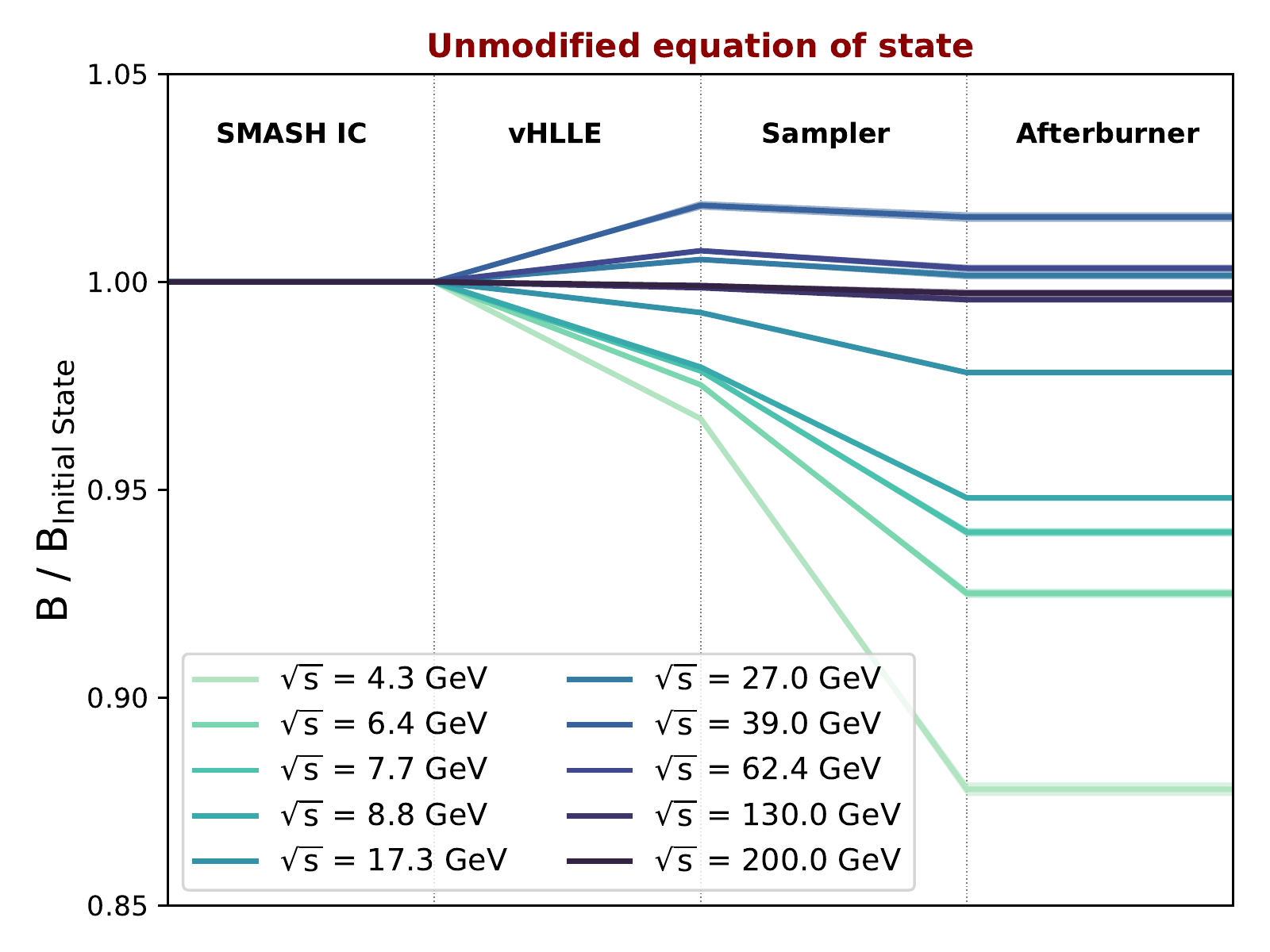}
  \hfill
  \includegraphics[width=0.29\textwidth]{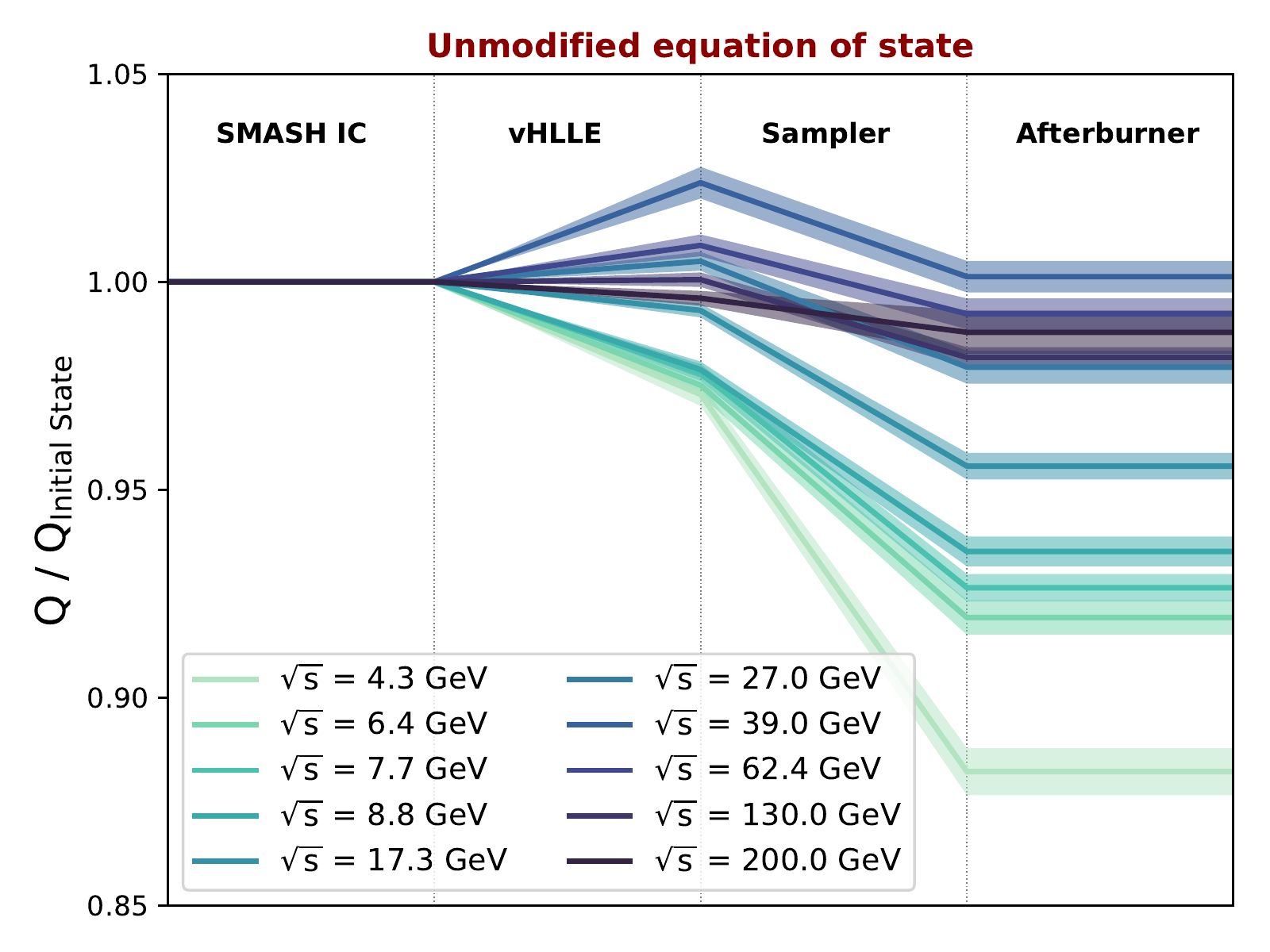}
  \\[+0.5cm]
  \includegraphics[width=0.29\textwidth]{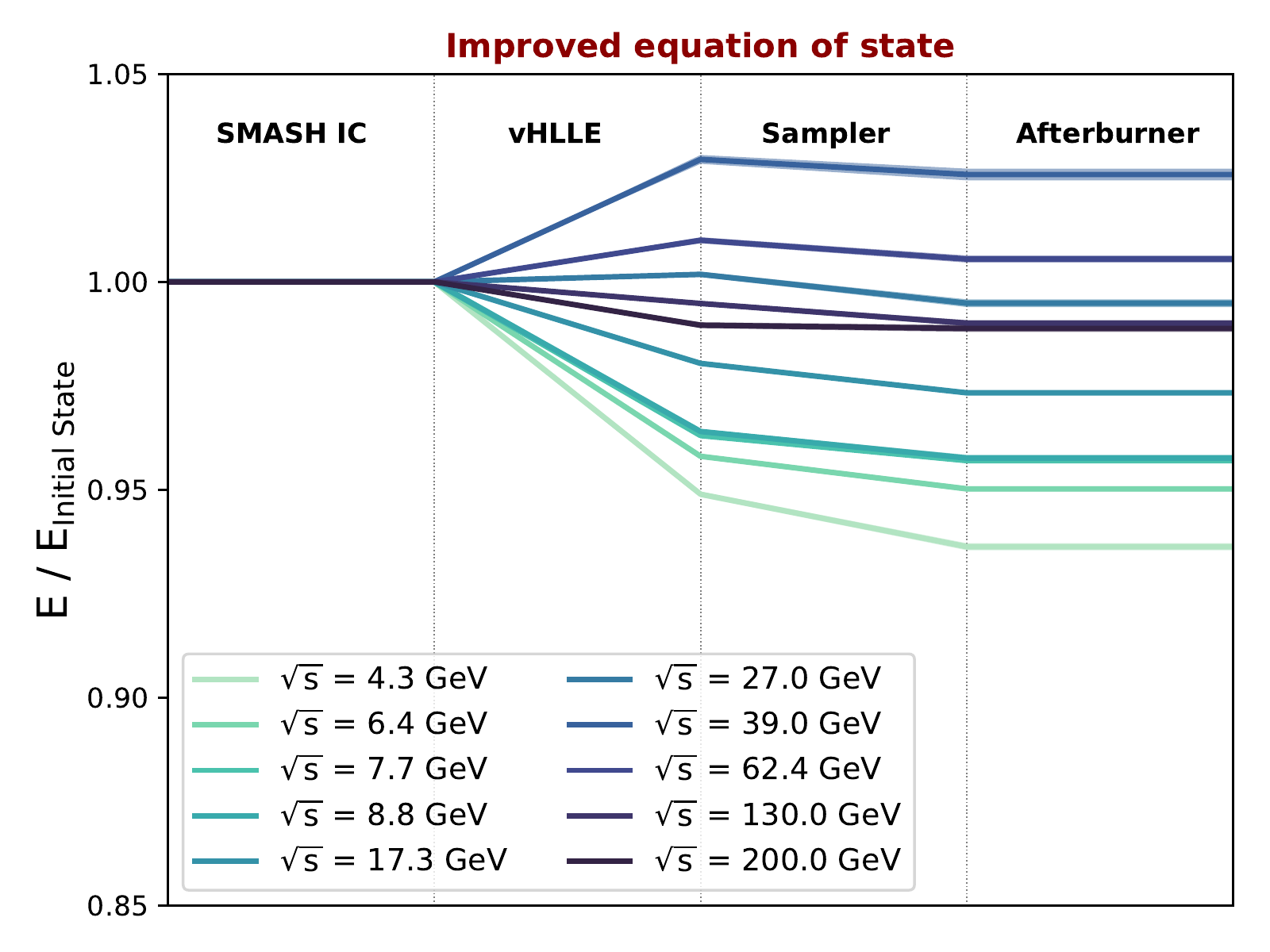}
  \hfill
  \includegraphics[width=0.29\textwidth]{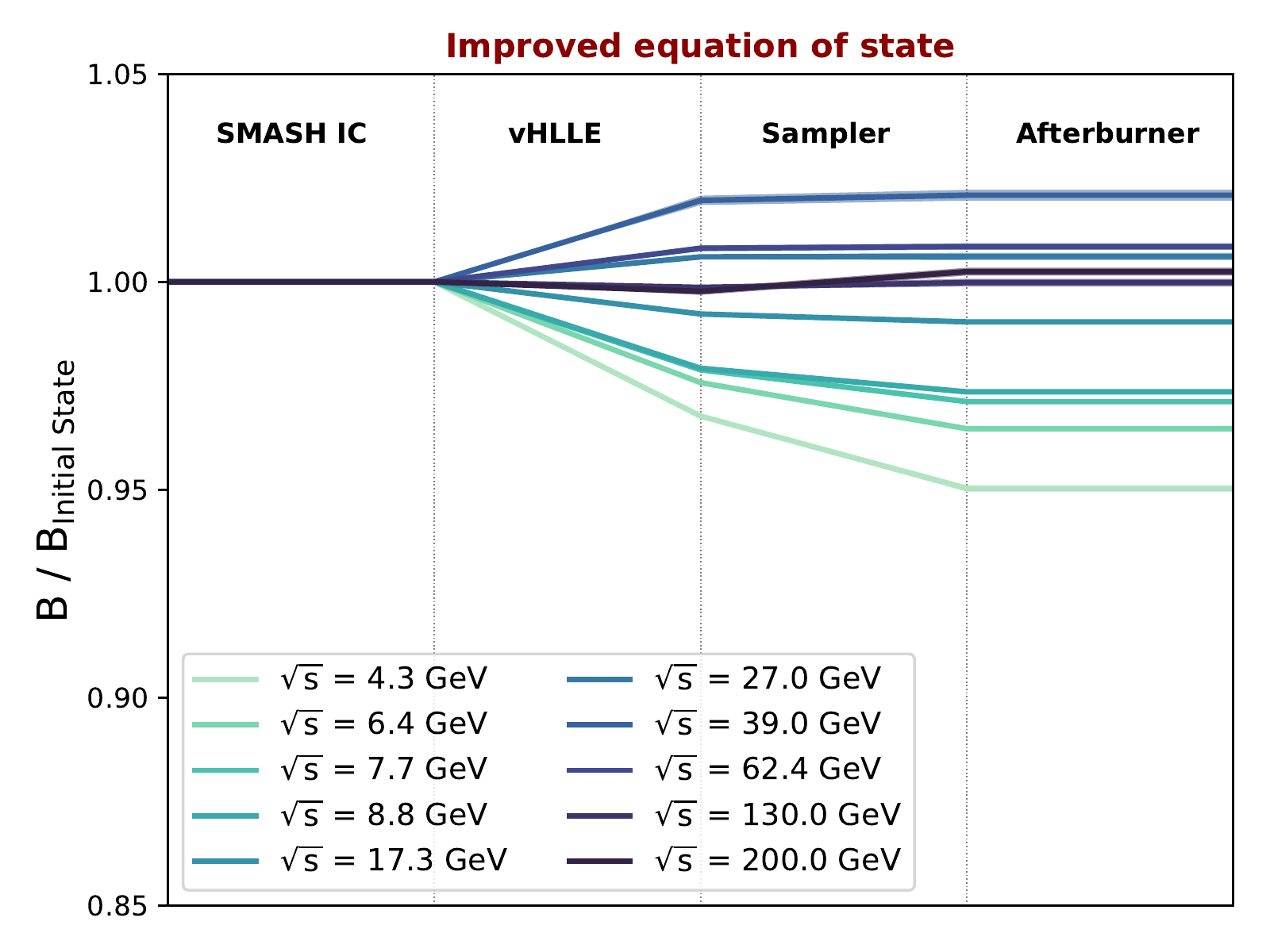}
  \hfill
  \includegraphics[width=0.29\textwidth]{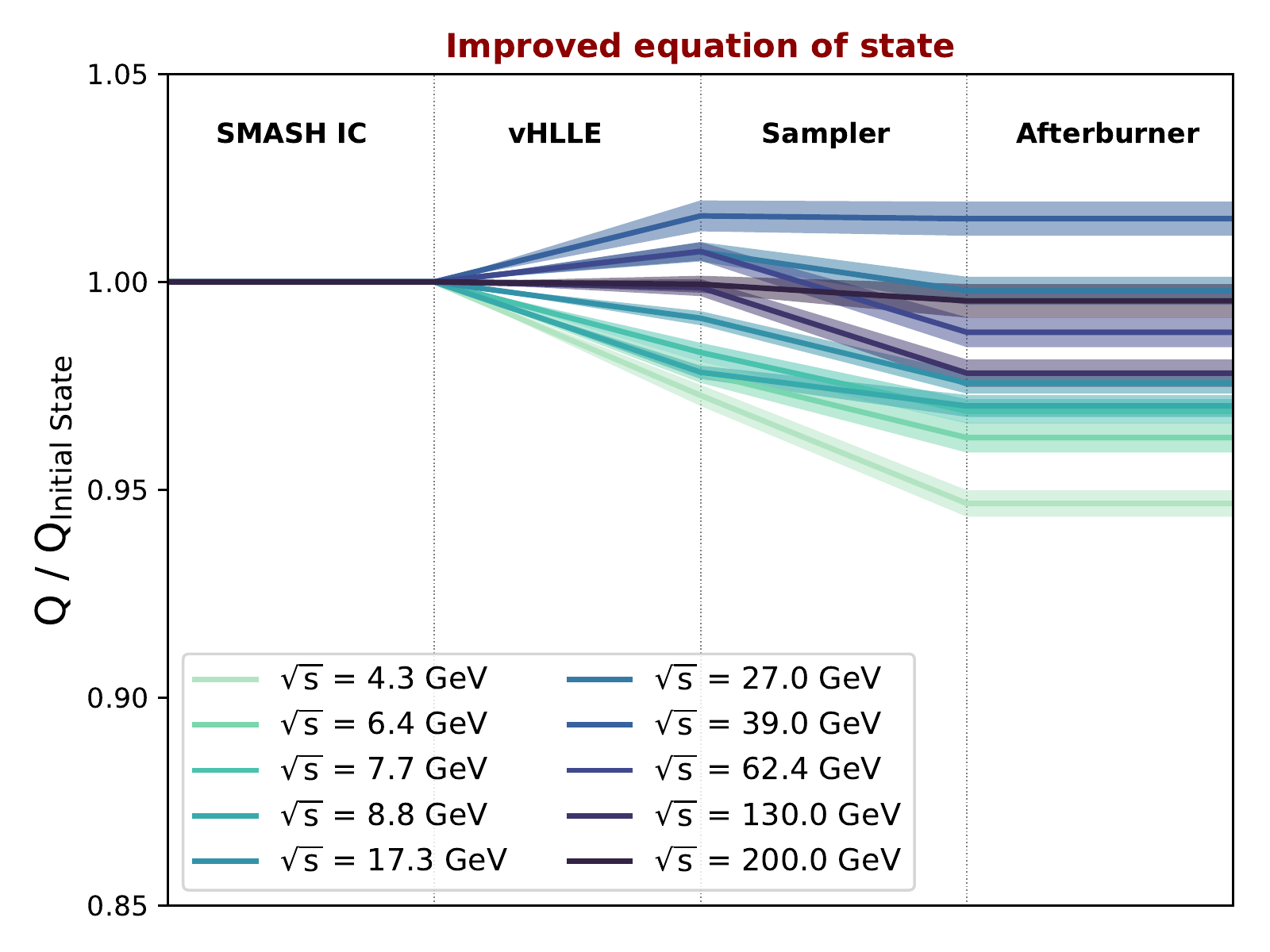}
  \caption{Conservation of total energy (left), baryon number (center) and electric charge (right) throughout all stages of the \texttt{SMASH-vHLLE-hybrid}. The leftmost ends corresponds to the initial state, the rightmost ends to the final state. The upper row of plots originates from a setup with the \textit{unmodified equation of state}, while for the lower row of plots the \textit{improved equation of state} of the \smsh hadron resonance gas was used. See Sec.~\ref{sec:model_eos} for further details about these equations of state.}
  \label{fig:Conservation}
\end{figure}

\begin{figure}[t]
  \centering
  \includegraphics[width=0.47\textwidth]{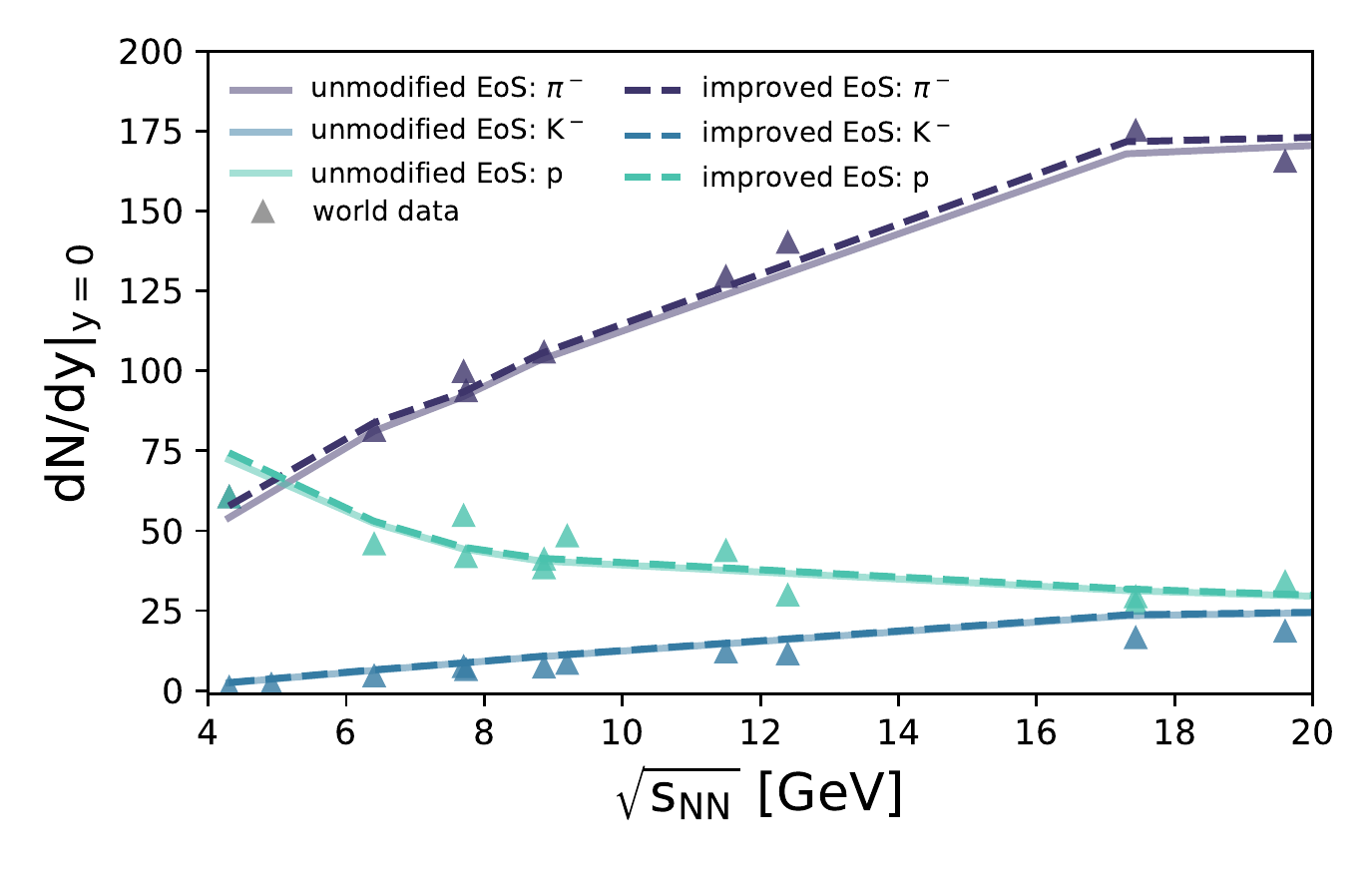}
  \hfill
  \includegraphics[width=0.47\textwidth]{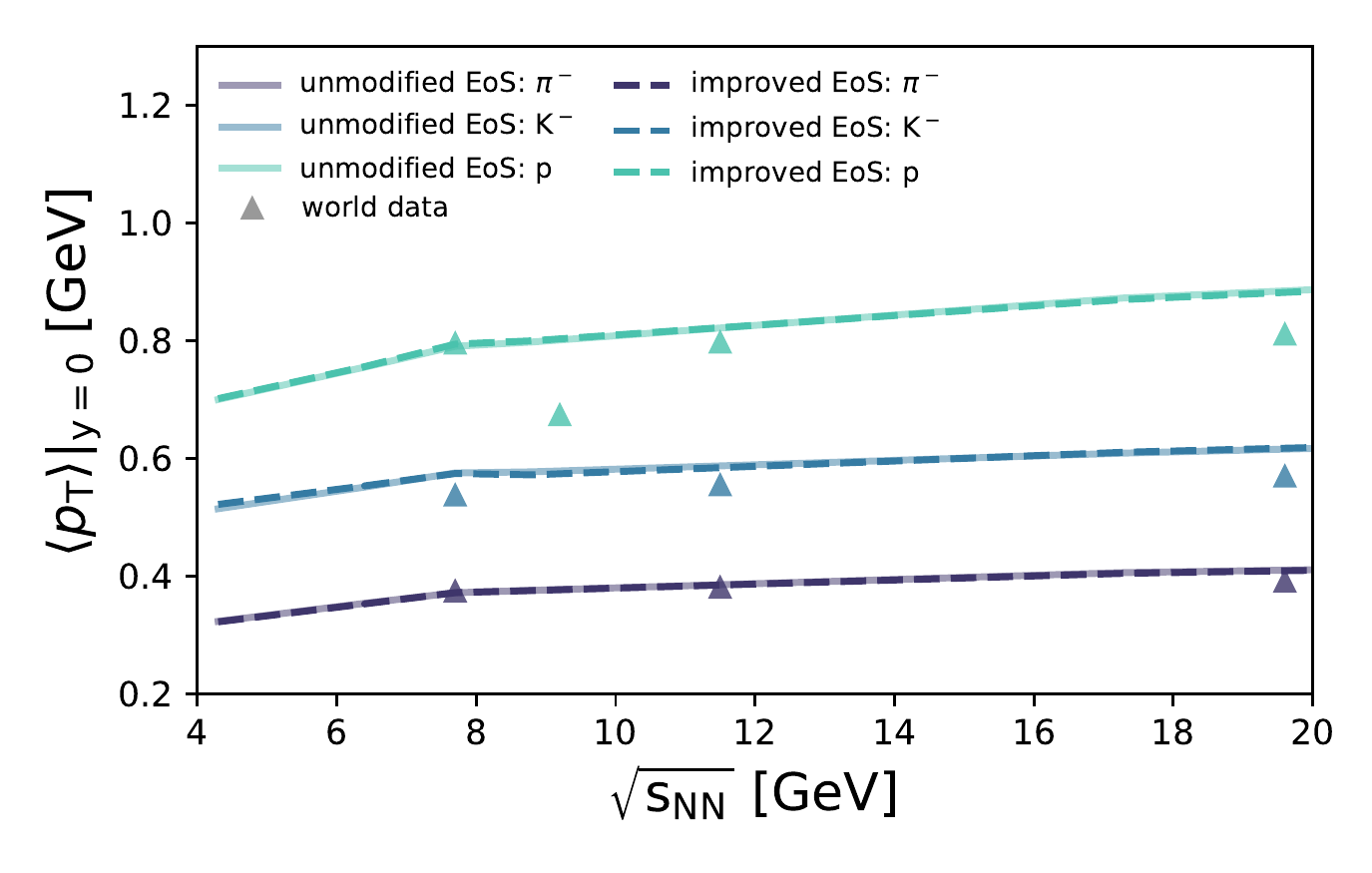}
  \caption{Midrapidity yield (left) and mean \pt (right) excitation function of \pim, $p$ and \Km, as obtained within the \hybrid utilizing the \textit{unmodified equation of state} (solid lines) and the \textit{improved equation of state} (dashed lines). The experimental data is collected from \cite{exp_data}.}
  \label{fig:exc_func}
\end{figure}

In what follows, first results from the \hybrid are presented in terms of excitation functions. First however, particular emphasis is laid on the conservation of quantum numbers $E$ (energy), $B$ (baryon number) and $Q$ (electric charge) throughout all stages of the collision, i.e. all modules embedded in the hybrid. Note that, in contrast to e.g. \cite{Oliinychenko:2020cmr, Vovchenko:2021yen}, the herein presented results correspond to global and on average conservation of $E, B$, and $Q$. To emphasize the importance of a well matching equation of state underlying the creation of the hydrodynamic freezeout hypersurface, the \hybrid is applied in two setups utilizing (i) the \emph{unmodified equation of state} and (ii) the \emph{improved equation of state} of the \smsh hadron resonance gas.
Although viscosities play an important role in hybrid models \cite{Karpenko:2015xea}, we decided to restrict this quantum number analysis to an ideal hydrodynamics setup.\\
In Figure~\ref{fig:Conservation}, the evolution of the conserved quantities $E$ (left), $B$ (center), and $Q$ (right) in the \hybrid are presented for heavy-ion collisions ranging from \sqrts = 4.3 GeV to \sqrts = 200.0 GeV.
The upper row corresponds to results obtained with the \emph{unmodified equation of state} and the lower row to those obtained with the \emph{improved equation of state}.
The subplots in Fig.~\ref{fig:Conservation} display the total quantum number ($E$, $B$ or $Q$) in each stage of the hybrid model, normalized to its respective initial value.
It can be observed that in the initial stage all quantum numbers are perfectly conserved, while small violations of conservation laws become apparent in the hydrodynamic stage. They are however found to be of the order of $< 6$ \% in all cases and are known to stem from (i) the transition to Milne coordinates, resulting in the time-integration accuracy of the associated source terms in each timestep to be finite, as well as from (ii) finite grid effects in the creation of the particlization surface, since the grid size of the underlying equation of state is also finite.
In the next stage, the sampling process, the effects of a mismatching hadronic equation of state become apparent. While for the \textit{unmodified equation of state} severe violations of energy, baryon number and charge conservation of up to 15\% are observerd, those quantum numbers are approximately conserved with the \textit{improved equation of state}. Note that quantum number violations in case of the \textit{unmodified equation of state} get more severe for lower collision energies. This is related to the fact that the \textit{unmodified equation of state} is especially troublesome at low energy densities. The lower the collision energy, the larger the fraction of cells falling into the problematic region, such that their relative contribution to the total quantum numbers $E, B$ and $Q$ is higher. In the afterburner stage all quantum numbers are again perfectly conserved, as quantum number conservation is enforced in \smsh. Summarizing the findings in Fig.~\ref{fig:Conservation}, a well matching hadronic equation of state at the particlization interface is of fundamental importance to fulfil conservation laws. Relying on the \textit{improved equation of state}, the latter are violated by no more than 7\% in the \hybrid across a large range of collision energies, which is considered an important validation of the model.
\\
To further demonstrate the impact of a mismatching equation of state on final state observables, the midrapidity yield and mean \pt excitation functions of \pim, $p$ and \Km, up to \sqrts = 20 GeV, are presented in Fig.~\ref{fig:exc_func}. Note that, in contrast to the previous study, the shear viscosities listed in Table~\ref{tab:parameters} are applied in the hydrodynamic evolution.
It can be observed that the midrapidity yields are noticeably increased once relying on the \textit{improved equation of state}, while the mean transverse momenta are insensitive. The increase of the former can be quantified to up to 8\% at low collision energies, but decreases as collision energies rise. This is in line with above observations of a matching equation of state being of greater importance for collisions at lower energies.
It shall further be noted that the resulting excitation functions for midrapidity yields as well as mean transverse momenta are in good agreement with experimental data, thus further validating the presented approach.

\section{Conclusions}
\label{sec:conclusions}
In this work, a novel hybrid model, the \texttt{SMASH-vHLLE-hybrid}, was introduced. It is suitable to describe heavy-ion collisions ranging from \sqrts = 4.3 GeV to \sqrts = 5.02 TeV. Therewith, we pointed out the fundamental importance of a well matching equation of state in the particlization process to ensure quantum number conservation. We demonstrated that energy, baryon number, and electric charge conservation are violated by no more than 7\% in an ideal hydrodynamics setup of the \texttt{SMASH-vHLLE-hybrid}, which is related to finite grid effects in the hydrodynamic stage upon transition to Milne coordinates and creation of the particlization surface.
In addition, multiplicity and mean \pt excitation functions were presented for \pim, $p$, and $K^-$, where only the former was sensitive to the choice of the equation of state, but both were in decent agreement with experimental data. \\
These results provide a first validation of the novel \hybrid at low and intermediate collision energies. In continuation, it can be applied to
a broader range of energies and observables to subsequently be confronted with experimental data. Furthermore, the \texttt{SMASH-vHLLE-} \texttt{hybrid} can be extended by more dynamical initial conditions, as for example realized in \cite{Akamatsu:2018olk}, to better capture the underlying dynamics of low-energy collisions.

\section{Acknowledgements}
A.S. acknowledges support by the Stiftung Polytechnische Gesellschaft Frankfurt am Main as well as the GSI F\&E program.
I.K. acknwowledges support by the Ministry of Education, Youth and Sports of the Czech Republic under grant “International Mobility of Researchers – MSCA IF IV at CTU in Prague” No. CZ.02.2.69/0.0/0.0/20 079/0017983.
Computational resources have been provided by the \mbox{GreenCube} at GSI.

%#############################
%
% Bibliography
%
%#############################
\bibliographystyle{rsc}

\end{document}